\documentclass[12pt]{article}
\usepackage{amssymb,amsmath,esint,titlesec}
\titleformat*{\section}{\normalsize\bf}
\titleformat*{\subsection}{\small\bf}

\begin{document}


\begin{titlepage}

\setlength{\baselineskip}{18pt}

                               \vspace*{0mm}

                             \begin{center}

{\Huge\bf Systolic aspects of black hole entropy}

                            \vspace*{3.5cm}

              \Large\sf  NIKOLAOS  \   KALOGEROPOULOS $^\dagger$\\

                            \vspace{0.2cm}
                            
 \normalsize\sf    Department of Mathematics and Natural Sciences,\\
                           The American University of Iraq, Sulaimani,\\
                           Kirkuk Main Road, Sulaimani,\\
                              Kurdistan Region, Iraq.                     \\

                            \end{center}

                            \vspace{3.5cm}

                     \centerline{\normalsize\bf Abstract}
                     
                           \vspace{3mm}
                     
\normalsize\rm\setlength{\baselineskip}{18pt} 

We attempt to provide a mesoscopic treatment of the origin of black hole entropy in (3+1)-dimensional 
spacetimes.  We ascribe this entropy to the
non-trivial topology of the space-like sections $\Sigma$ of the horizon. 
 This is not forbidden by topological censorship,
 since all  the known  energy inequalities needed to prove the spherical topology of $\Sigma$ are violated in quantum theory.
We choose the systoles of $\Sigma$ to encode its complexity, which gives rise to the black hole  entropy. 
 We present hand-waving reasons  why  the entropy of the black hole can be considered as a function of the volume entropy of $\Sigma$. 
We  focus on the limiting case of $\Sigma$ having a large genus.  \\

                           \vfill

                                                                         
                             \vfill

\noindent\rule{9cm}{0.2mm}\\  
   \noindent $^\dagger$ \small\rm Electronic address: \ \  \  {\normalsize\sf   nikos.physikos@gmail.com}\\

\end{titlepage}
 

                                                                                \newpage                 

\rm\normalsize
\setlength{\baselineskip}{18pt}

\section{Introduction}

The statistical origin of the black hole entropy \cite{Sorkin1, Wald, Carlip, FF, JacP} has been a perplexing problem since the earliest works on black hole thermodynamics 
more than forty years ago \cite{Bek1, Bek2, Hawk1}. Numerous proposals have been put forth over these years, about such an origin, 
all of which have their strong and weak points. As an example of such proposals, which has attracted considerable attention  recently, 
is the entanglement entropy \cite{Solod, Nish} (and references therein). Candidates for a quantum theory of gravity, such as loop gravity \cite{Perez}
causal sets \cite{SorYaz}, and String / M-theory \cite{RT},  face as an important test 
the microscopic origin of black hole /``horizon" entropy, the derivation of the Bekenstein and Hawking formulae and their possible extension, from their assumptions and 
within their corresponding frameworks.  \\

In this work we address a semi-classical/mesoscopic aspect of this problem. Classically, a space-like section \ $\Sigma$ \ of the horizon
whose area \ $A(\Sigma)$ \ is a multiple of the entropy,
 must have spherical topology: this is the content of a theorem due to Hawking \cite{Hawk2, HE}, and of the closely related ``topological censorship". 
 Hawking's theorem applies to vacuum asymptotically flat stationary spacetimes in (3+1) dimensions. Beyond this point, one can naturally ask what happens 
 in more realistic situations when there is some form of matter/energy present in such spacetimes.\\
 
To address this, Hawking's theorem and the subsequent topological censorship theorems assume the validity of some classical energy condition on the stress-energy tensor, 
namely a lower bound on its contraction with time-like or causal vectors \cite{HE}.  
The most frequently invoked such energy conditions are the weak energy and the dominant energy conditions, both of which appear to be reasonable at the classical level. 
However, all classical energy conditions such as the ones in \cite{HE} are known to be violated, point-wise at least, at the quantum level. 
This allows for the possibility of the sections \ $\Sigma$ \ of the horizon of such stationary asymptotically flat spacetimes even in (3+1)-dimensions
 to have any topology, at least at such a mesoscopic level. One can allow for such a possibility in fully dynamical spacetimes, where even the definition 
 of what constitutes a black hole horizon is still subject to speculation and controversy \cite{Faraoni}.\\

 To address this point, we use a ``semi-phenomenological" approach. We are largely motivated by the fact that the horizon may have some ``wrinkling'' \cite{Sorkin2, CFGH},
which can be used as the source of the black hole entropy. We push this fact to a logical limit, in the present work,  by assuming that this ``wrinkling" is a topology change 
of the horizon which is not precluded because of the violation due to quantum effects  of the clasiscal energy conditions.  
 So, we explore the option of ascribing the entropy of  black holes to the possible mesoscopically non-trivial topology of \ $\Sigma$. \ 
 We use, in particular,  the length of the smallest non-contractible loop on such a surface as a linear measure characterizing the elementary 
 ``cells" that the black hole entropy counts.\\

  The advantage of such a ``systolic" approach  is that it does not depend on any curvature bounds of the 
 induced Riemannian metric on \ $\Sigma$. \ This is  desirable because through Einstein's equation, a Ricci curvature bound amounts to a bound 
 on the stress-energy tensor, its powers, derivatives and contractions, for which very little is generically known even semi-classically. 
 We choose the volume entropy of \ $\Sigma$ \  to encode the ``lack of order"/``complexity" giving rise to the statistical entropy of the black hole.  
 We point out relations between the volume entropy and the systoles of \ $\Sigma$ \ paying attention to the particular case of such
 \  $\Sigma$ \ having a large genus \ $g$. \ In a very closely related work \cite{NK1}, we had explored the possibility of assuming the preservation of  the spherical topology of \ $\Sigma$, \ where we ascribed the thermodynamic entropy of \ $\Sigma$ \ to a different measure of its complexity  rather than its genus, which in the case of   
spherical topology  is trivial. \\
 
In Section 2, we provide some background  about (homotopic) systoles, in order to make our presentation somewhat self-contained. 
Section 3 presents the core of our proposal, arguing about the use of the volume entropy and stating bounds involving it, and systolic data, providing 
physical motivation and interpretation wherever possible. Section 4 presents some conclusions and a brief outlook toward the future. \\


\section{Some background on systolic geometry}

The (homotopy) 1-systole \  $Sys (\mathcal{M}, \mathbf{g})$ \ of a Riemannian manifold \ $(\mathcal{M}, \mathbf{g})$ \ is the infimum of the lengths of non-contractible 
loops of that manifold. In symbols, let \ $\gamma: \mathbb{S}^1 \rightarrow \mathcal{M}$ \ be a closed loop of length \ $L(\gamma)$ \ on \ $\mathcal{M}$ \ which is not 
contractible. Then 
\begin{equation}
         Sys (\mathcal{M}, \mathbf{g}) \ = \ \inf_\gamma L(\gamma) 
\end{equation}   
This clearly assumes that \ $\pi_1(\mathcal{M})$ \ is non-trivial, hence the 2-dimensional surfaces under consideration in this work have to have genus \ $g\geq 1$, \
which excludes the case of the 2-sphere \ $\mathbb{S}^2$ \ since \ $\pi_1(\mathbb{S}^2)$ \ is trivial. For any compact Riemannian manifold 
the infimum is attained, and the loop realizing it is a simple closed geodesic.\\
 
\noindent C. Loewner \cite{Pu} proved for the 2-dimensional torus \ $(\mathbb{T}^2, \mathbf{g})$ \  the systolic inequality
\begin{equation} 
        \frac{2}{\sqrt{3}}  \ A(\mathbb{T}^2, \mathbf{g}) \ \geq \  [Sys (\mathbb{T}^2, \mathbf{g})]^2
\end{equation}
The equality is attained by the flat equilateral metric on \ $\mathbb{T}^2$. \  This metric is homothetic to a quotient of \ $\mathbb{C}$ \ by the lattice spanned by 
the cubic roots of unit. The constant \ $w_2 = 2/\sqrt{3}$ \  is the Hermite constant defined by 
\begin{equation}
      w_2 \ = \ \sup_{\Lambda \in \mathbb{R}^2} \left\{  \frac{\lambda_1(\Lambda)}{\sqrt{ A (\mathbb{R}^2 / \Lambda)}}  \right\}^2
\end{equation}
where the supremum is taken over all lattices \ $\Lambda$ \ in \ $\mathbb{R}^2$ \  endowed with the Euclidean distance function and its induced norm. 
Here \ $\lambda_1 (\Lambda)$ \  indicates the least length of a non-zero vector in \ $\Lambda$ \ which is considered as a lattice of \  $\mathbb{R}^2$ \ as noticed above. \\

\noindent  Subsequently, P. Pu \cite{Pu} proved for the projective plane \ 
$(\mathbb{RP}^2, \mathbf{g})$  \  that
\begin{equation}
        \frac{\pi}{2} \ A(\mathbb{RP}^2, \mathbf{g}) \ \geq \  [Sys (\mathbb{RP}^2, \mathbf{g})]^2 
\end{equation}
The equality in this case is attained by the metric of constant Gaussian curvature. We recall at this point that the Gaussian curvature is the only fundamental 
curvature invariant (non-trivial part of the Riemann tensor) of  a surface. Hence the metric saturating Pu's inequality if that of the round sphere with antipodal 
points identified. At this point one would  probably be amiss not to quote the early works \cite{Acc, Blatter} on systoles. For a pedestrian introduction to systolic 
geometry which includes the best known constants, at the time of its writing, one can consult \cite{Berger}.\\

If one defines the systolic ratio of a surface \ ($\Sigma, \mathbf{g}$) \ as
\begin{equation}
    \mathsf{SR} (\Sigma, \mathbf{g}) \ \equiv \  \frac{[Sys (\Sigma, \mathbf{g})]^2}{A (\Sigma, \mathbf{g})}
\end{equation}
 then Loewner's inequality (2) can be re-written as 
 \begin{equation}
     \mathsf{SR} (\mathbb{T}^2, \mathbf{g}) \ \leq \ \frac{2}{\sqrt{3}}
 \end{equation}
 and Pu's inequality (4) as
 \begin{equation}
        \mathsf{SR} (\mathbb{RP}^2, \mathbf{g}) \ \leq \ \frac{\pi}{2} 
 \end{equation}
Other results in the above spirit exist  for the  Klein bottle $(\mathbb{K}, \mathbf{g})$ \cite{Blatter, Bavard}. The systolic ratio is bounded above  
\begin{equation}
      \mathsf{SR} (\mathbb{K}, \mathbf{g}) \ \leq \ \frac{\pi}{2\sqrt{2}}
\end{equation}
Perhaps counter-intuitively compared to (2), (4), the equality in (8) turns out to be realized by a metric produced by gluing two copies of the M\"{o}bius 
strip along a singular line, each M\"{o}bius strip being obtained from the standard sphere.   In general, one cannot hope to find the metrics, 
or the values of the upper or lower bounds on the systolic ratio. Usually one has to settle with less: either finding some bounds of the systolic ratio or, at 
best, some bounds for given genus $g$ surface,  or bounds for a large genus ($g\rightarrow\infty$) surface. \\  

Since we will only be interested in 2-dimensional surfaces with their induced metrics \  ($\Sigma, \mathbf{g}_\Sigma$) \  which are space-like sections of 
horizons of (3+1)-dimensional spacetimes  ($\mathcal{M}, \mathbf{g}$), we will forego any discussion of the systoles of higher dimensional space-times, 
of higher dimensional systoles,  of homological systoles,  and of various related issues, and refer the interested reader to some of the original and expository references 
\cite{Berger, Gromov1, Gromov2, Gr-book, Guth, Katz-book} on these matters.\\

We will focus on the asymptotic case, where the genus \ $g\rightarrow\infty$. \ The physical reasons motivating such a choice will be explained in the next section.    
An upper bound for the systolic ratio for large genus surfaces that depends on the genus was given in \cite{Gromov1}. A lower bound was given by \cite{BuSar, KSV}.
Other notable contributions (the list is non-exhaustive) in this direction are \cite{KS1, KS2, Katz-book}. 
What is pertinent for this work is that for large genus  \ $(g\rightarrow\infty$) \ the systolic ratio of a surface \ ($\Sigma,\mathbf{g}$) \  behaves as 
\begin{equation}
     \mathsf{SR} (\Sigma, \mathbf{g}) \ \sim \  \frac{(\log g)^2}{g}
\end{equation}
where the symbol  \ ``$\sim$" \ means here that there are two positive constants (independent of \ $g$) \ $c_1, c_2$ \ such that
\begin{equation} 
          c_1  \ \frac{(\log g)^2}{g}  \ \leq \ \mathsf{SR} (\Sigma, \mathbf{g}) \ \leq \ c_2 \   \frac{(\log g)^2}{g}
\end{equation}
The significance of (10), and the main motivating factor for looking at systoles as a way of mesoscopically encoding the entropy of black holes, is that is it 
curvature-free. A large body of work in Riemannian geometry \cite{Berger} is about establishing inequalities between geometric quantities which are valid, as long as some 
of the curvature-related quantities of such manifolds have an upper or lower bound. However, this is problematic when it comes to semi-classical or quantum gravity as 
will be explained in the next Section. For this reason, curvature-free inequalities such as (10), can be very desirable in determining  mesoscopic features of gravitational 
systems  without actually needing to delve deeper, at this stage, into a quantum theory of gravity, which  should supposedly resolve, in principle at least,  
all such questions.\\        


\section{Systoles: geometry, topology and entropy}
 

\subsection{Horizons and energy inequalites}
 
 The usual (zero cosmological constant) Einstein field equations on a (3+1)-dimensional spacetime \ $(\mathcal{M}, \mathbf{g})$ 
\begin{equation}
     R_{\mu\nu} - \frac{1}{2} \ R \  \mathbf{g}_{\mu\nu} \ = \  \frac{8\pi G}{c^4} T_{\mu\nu}
\end{equation}
where \ $\mu, \nu = 0, \ldots, 3$ \ and where \ $R$ \ stands for the scalar curvature,  can be re-written as 
\begin{equation}
     R_{\mu\nu} \ = \ \frac{8\pi G}{c^4}  (T_{\mu\nu} - \frac{1}{2} T  \mathbf{g}_{\mu\nu} )
\end{equation}
where \ $T$ \ stands for the trace of the stress-energy tensor \ $T_{\mu\nu}$. \ These equations are identities, devoid 
of any physical meaning, unless
someone can put some constraints on the properties of \ $T_{\mu\nu}$. \ If not, then any Lorentzian-signature metric
(compatible with the topology of $\mathcal{M}$) would be a solution to the Einstein equations. In order to establish that the 
formation of singularities is not a specific feature of the Schwarzschild solution and due to its high degree of symmetry, 
but a generic feature  of General Relativity, the classical energy conditions were formulated during the 1960s \cite{HE, MMV}.   
These are ``reasonable assumptions" regarding the properties of the stress-energy tensor and they seem to be valid at the classical 
level. However they are all violated at the quantum level, a fact that has motivated the introduction of averaged, rather than point-wise
energy conditions or ``quantum inequalities" \cite{MMV, LF, CF}. These energy conditions having lower bounds seem 
to be obeyed in some particular cases, even though no definitive proofs exist generally. \\

The crux of problem is that there is no known, a priori, way of distinguishing between 
what are ``reasonable" and ``unreasonable" stress-energy tensor properties, in  order to be able to exclude, even by hand, the 
unreasonable ones from further consideration. Such a need becomes particularly acute today on the face of 
the accelerating expansion of the Universe and its conjectured solution via ``dark energy". Whether such a physical entity exists and if so,
 what are its properties \cite{DES1} is a subject of much contention theoretically and observationally. Therefore,
determining the bounds of its stress-energy tensor may also have immediate  observational consequences, apart from its general 
theoretical interest.   \\

Even if were able to put some bounds on the properties of the stress-energy tensor, its quantization would bring back some of these 
undesirable properties, which seem to be generically inescapable. Therefore it is very hard to come up with reasonable bounds 
on the stress energy tensor which would constrain the range of acceptable solutions to the Einstein field equations even in quantum 
field theory on curved spacetime, let alone in quantum gravity.  \\  

Things only become harder if one wishes to define what a black hole horizon is \cite{HE, Faraoni, AK, Booth, GJ, Chr}. 
The event horizon is very robust, but global and non-teleological. The quasi-local definitions are either kinematical (not taking into account 
the Einstein equations) or  involve projections of the field equations in codimension-2 surfaces \ $\Sigma$, \ which result in expressions 
that are hard to control analytically, and from which to extract useful physical information, without several additional simplifying assumptions. 
Due to this state of affairs, having curvature-free estimates, such as the systolic expressions, pertinent to black hole horizons may be advantageous in a way, 
even though they are only loosely related to  the underlying dynamics arising via the projection of Einstein's equations onto \ $\Sigma$.\\

If the use of  event horizons is not desirable, someone could use instead  ``marginally outer trapped surfaces" \cite{Gal-Lect}  which in the case of stationary black holes coincide
with the event horizons. Since we will only refer to situations close to equilibrium, the use of marginally outer trapped surfaces would not, naively at least,
give results to different from the ones that event horizons might. Considering only quasi-equilibrium situations is partly motivated by that very little is known
about the statistical mechanics of systems far from equilibrium. In particular, it is not even clear that a state function such as the entropy can be meaningfully defined or what its 
definition might be and, if so, what would be its precise physical interpretation  in the thermodynamic description of systems far from equilibrium.\\   

 
 \subsection{Systolic area}
 
The viewpoint that we adopt in the present work is that the origin of the black hole entropy depends on the genus \ $g$ \ of the space-like section \
($\Sigma, \mathbf{g}_\Sigma$) \ of the horizon. An advantage is that the genus is a  diffeomorphism invariant concept. Even more so, it is an invariant up 
to homotopy equivalence, a fact which makes it very robust.  The length of the systole can be seen as providing the linear measure of the minimum topological scale that 
can be used to measure the non-triviality of the entropy of \ $\Sigma$. \ As a result, the systolic ratio \ $\mathsf{SR}(\Sigma, \mathbf{g}_\Sigma)$ \ is the quantity 
that appears to be the most relevant for entropy, or to be more precise its inverse, which is called the systolic area  
\begin{equation}
    \sigma (\Sigma, \mathbf{g}_\Sigma) \ = \ [\mathsf{SR}(\Sigma, \mathbf{g}_\Sigma)]^{-1}
\end{equation}
Substituting (13) into (9) we get
\begin{equation}
      \sigma (\Sigma, \mathbf{g}_\Sigma) \ \sim \frac{g}{(\log g)^2}, \hspace{10mm} g\rightarrow \infty
\end{equation}
There is a natural lower bound for the length of a systole: the Planck length \ $l_P$. \ 
Below this scale, the classical geometry is not expected to describe nature accurately, and a quantum description is required. Hence
\begin{equation} 
    \sigma (\Sigma, \mathbf{g}_\Sigma) \ \leq \ \frac{A(\Sigma, \mathbf{g}_\Sigma)}{l_P^2}
\end{equation}  
and this is what someone should have in mind in taking the limit \ $g\rightarrow\infty$. \  However, since \ $A(\Sigma, \mathbf{g}_\Sigma)$ \ is macroscopic, the ratio in 
the right hand side  of (15) is of the order of magnitude of \ $10^{72}$ \ for a  solar mass Schwarzschild black hole.  Therefore taking the limit \ $g\rightarrow\infty$ \ 
is not too different from \ $g$ \ having as an upper bound
\begin{equation}   
         \frac{g}{(\log g)^2} \   \sim \ \frac{A(\Sigma, \mathbf{g}_\Sigma)}{l_P^2}
\end{equation}

\vspace{3mm}


\subsection{On the topology of space-like sections of  horizons}

The fact that the horizon of a black hole should have space-like sections \ $\Sigma$ \ of arbitrarily large genus \ $g$, \  seems to contradict the conclusion of  
a theorem and its extensions, initially due to Hawking, which only allows for the topology of such sections to be spherical \cite{Hawk2, CW}. 
There is also the very closely related   
issue of ``topological censorship" stating that a spacetime cannot have any observable topologically non-trivial features, as such features would 
collapse too quickly to allow someone to detect them \cite{FSW}. Hawking's theorem and issue of topological censorship have been revisited under 
weaker \cite{JV}, modified assumptions \cite{Gal, GSWW1}, in higher dimensions \cite{GalSch, KWW}, in dS/CFT and AdS/CFT correspondences \cite{GSWW2,  
AG}, and in modified theories of gravity of potential astrophysical interest \cite{BamMod}, to just name few cases \cite{FriedHig, ADGP}. \\

The common denominator 
in such cases are the assumptions of having a $(3+1)$-dimensional spacetime, and also having asymptotic flatness, except where explicitly stated otherwise. 
Most important is the assumption of the validity of the dominant or the weak energy conditions. However, as stated above, all the classical energy conditions are violated   
quantum mechanically, hence it may not be unreasonable to assume that no constraint on the genus \ $g$ \ of a space-like section of the horizon can be placed 
 in the context of quantum field theory on a black hole space-time. As a result, we are compelled to examine the case of \ $\Sigma$ \ being a higher genus \ $g$ \  
 surface and, more specifically, to focus on the extreme, asymptotic, case of \ $g\rightarrow \infty$. \\ 

A question that should be addressed in the current proposal is how  \ $\Sigma$ \ makes the transition from having a large genus \ $g$ 
\  in the semiclassical/quantum case, to having spherical topology \ ($g=0$) \ in the classical limit.  We have no answer to this question, 
only a conjectural plan of how things might possibly  work. A  surface of genus \ $g$ \ is topologically a sphere with \ $g$ \ handles attached to it. 
The conjecture is that in the classical limit these handles should progressively become thinner and thinner until their volume 
degenerates, in which case their contribution to entropy would progressively vanish.\\

  Assume that due to the spacetime symmetries and in the time symmetric case,  
or due to lack of them in the generic case, such a \ $\Sigma$ \ has a constant Gaussian curvature metric \ $\mathbf{g}|_\Sigma$, \  
which upon re-scaling by a constant factor, is hyperbolic. For  2-dimensional manifolds there is  a thin-thick decomposition \cite{Gr-book} which 
simplifies substantially in the  hyperbolic case. \\

Let the injectivity radius of \  $x\in\Sigma$ \  be indicated by \ $\mathsf{inj}_x(\Sigma)$ and let \  $c>0$ \ be  a positive constant. 
The thin part \ $\Sigma_{<c}$ \ of \ $\Sigma$ \ is the set of all points \ $x\in\Sigma$  \  such that \ $\mathsf{inj}_x(\Sigma) < c$. \ 
Its complement is the thick part  \ $\Sigma_{\geq c}$ \ of \ $\Sigma$. \ Then one trivially has 
\begin{equation}     
     \Sigma \ = \ \Sigma_{<c} \cup \Sigma_{\geq c}
\end{equation}
Let's assume that the black hole under study is in quasi-equilibrium. Then, in the simplest case, \ $\Sigma$ \ should be at least a locally homogeneous space. 
Due to the Gauss-Bonnet theorem for the closed 2-manifold \ $\Sigma$ \ of genus \ $g$    
\begin{equation}
     \frac{1}{2\pi} \int_\Sigma  R  \  dvol_\Sigma \ = \ \chi(\Sigma) 
\end{equation}
where \ $R$ \ stands for the Ricci scalar of the induced metric \ $\mathbf{g}_\Sigma$ \ and \  $\chi(\Sigma)$ \ is the Euler characteristic of \ $\Sigma$ \  
which turns out to be 
\begin{equation}
   \chi(\Sigma) \ = \ 2-2g
\end{equation}
  It is probably reasonable to expect that such a metric will not have any 
point-like (Dirac delta), linear or higher-dimensional simplicial curvature singularities, at least in stationary quasi-equilibrium cases. Any such simplicial 
curvature singularities would be unstable  and would eventually bring about the local homogeneity of \ $\Sigma$ \
based on the behavior of more conventional macroscopic systems when they are close to equilibrium. \\

Due to it local homogeneity, and for a large genus \ $g$, \ the scalar curvature of \ $\Sigma$, \ which coincides with the 
Gaussian curvature (as is always true for smooth 2-dimensional surfaces) would be negative everywhere and its value, therefore, 
would be bound away from zero, according to the Gauss-Bonnet theorem. The hyperbolic case, where 
the universal cover of \ $\Sigma$ \  is the hyperbolic plane \ $\mathbb{H}^2$ \ is, in some sense optimal, as it is the unique space form of negative 
sectional curvature, and is therefore used for general comparison purposes.\\

 One can prove that for a surface \ $\Sigma$ \ with a metric of negative sectional curvature, and for \ $c$ \ smaller than the Margulis constant of \ $\Sigma$ \ 
 \cite{Gr-book}, each component of \ $\Sigma_{<c}$ \ is either a cusp or an annulus. Cusps are unbounded, and as such they are not acceptable, on physical 
 grounds,  as parts of \ $\Sigma$. \  The annuli  are tubular neighborhoods of closed geodesics of \ $\Sigma$ \ of length smaller than \ $c$. \  
 These are bounded and diffeomorphic to a circle \ $\mathbb{S}^1$  \  times an interval.\\
 
 The inevitable existence  of quantum states of the stress energy tensor violating the classical 
energy inequalities would have as  a result the existence of annuli in \ $\Sigma_{<c}$. \  However,  in the classical limit, such states would be less and less 
probable or would contribute less and less in the partition function. As a result,  the area of such annuli will decrease, until they would
disappear altogether in the classical limit. This would signify that the circle of the annuli would get a continuously decreasing radius until it collapses to a point, 
and the annulus degenerates to a line segment. In the classical limit, this would make the horizon disconnected. Since in the definition of horizon one assumes that
it is connected, the disconnected parts will become thinner and thinner and progressively more and more irrelevant for the partition function.
The final remnant will be a topological sphere. 
In other words, if \  $\Sigma$ \ is seen as a topological sphere with \ $g$ \ handles attached, the above process amounts to each of these 
\ $g$ \ handles degenerating  in area.  The final remnant  is a topological sphere, which was initially  the thick  part of \ $\Sigma$.\\
 
 
\subsection{Entropy from the topology of \  $\Sigma$} 
 
Since the entropy is, in a way, a measure of the ``lack of order" or ``complexity" of a system, one might wonder how the entropy of such a space-like section  
\ $\Sigma$ \ of the horizon can be quantified. 
Because there is a minimum physical length, the Planck length \ $l_P$, \ below which the validity of General Relativity, or any classical gravitational theory, 
is  non-applicable, we have chosen to use 1-dimensional objects, such as the (homotopy) systoles to express such a ``lack of order", 
stemming from the topology of \ $\Sigma$. \ Since we lack a quantum theory of gravity which could guide us to look for a ``natural" measure of such a complexity, 
we have to make a judicious choice. Our quantity of choice will be a function of the  ``asymptotic volume", which is also known as ``volume entropy", 
\ $h_{vol}(\Sigma, \mathbf{g}_\Sigma)$ \ of \ $\Sigma$. \\

We have to state right away that  even though the latter name is highly suggestive, \ $h_{vol}$ \ is a purely geometric quantity which has nothing whatsoever to do with 
the physical Clausius or Boltzmann/Gibbs/Shannon (BGS) entropy \ $\mathcal{S}_{BGS}$, \ in general. Admittedly, the definition of \ $h_{vol}$ \ is motivated by and shares 
formal properties with the statistical definition \ $\mathcal{S}_{BGS}$ \ of entropy. 
 It also happens that \ $h_{vol}$ \ has been used in some models of possible physical interest as a 
substitute for \ $\mathcal{S}_{BGS}$. \ These are the analogies and aspects of \ $h_{vol}$ \ that we wish to explore in the present work. \\

In general \cite{KatHas}, consider a closed (compact, without boundary) Riemannian manifold  \  ($\mathcal{M}, \mathbf{g}$) \ and let \  ($\tilde{\mathcal{M}}, \tilde{\mathbf{g}}$) \  
be its universal cover. Let \ $\tilde{x}_0 \in\tilde{\mathcal{M}}$. \  Then \ $h_{vol}(\mathcal{M}, \mathbf{g})$ \  is defined as 
\begin{equation}        
 h_{vol} (\mathcal{M}, \mathbf{g}) \ = \ \lim_{r\rightarrow +\infty} \frac{\log (vol \ B(\tilde{x}_0, r))}{r}
\end{equation}
The volume on the right-hand-side of (20) is computed with the metric \ $\tilde{\mathbf{g}}$, \ and \ $B(x_0, r)$ \ indicates the ball centered at \ $\tilde{x}_0$ \ 
and having radius \ $r$ \ in \ $\tilde{\mathcal{M}}$. \ Since \ $\mathcal{M}$ \ is compact, the limit in (20) exists and it is independent of the choice of \ $x_0$. \ 
As one can immediately see \ $h_{vol} (\mathcal{M}, \mathbf{g})$ \ expresses the exponential growth rate of the volume of the universal cover.  Therefore it is 
particularly well-suited to give non-trivial results for manifolds of negative sectional curvature such as \ $\Sigma$.  \\

The quantitative property that makes the volume entropy appealing for our purposes, is that for a closed Riemannian manifold \ ($\mathcal{M}, \mathbf{g}$) \ 
the following holds \cite{KatHas}: let \ $y_0\in\mathcal{M}$ \ and let \ $\mathcal{N}(s)$ \ be the number of homotopy classes of loops based at \ $y_0$ \  which have loops of 
length at most \ $s$. \ Then 
\begin{equation}  
    h_{vol}(\mathcal{M}, \mathbf{g}) \ = \ \lim_{s\rightarrow +\infty} \frac{ \log \mathcal{N}(s)}{s} 
\end{equation} 
Therefore, in a particular sense, the volume entropy of \ $\Sigma$ \ expresses the exponential ``lack of order"/``complexity" of \ $\Sigma$ \ 
from a topological (homotopic) viewpoint, as it is probed by 1-dimensional objects (loops). \\

One could wonder about the reason why we do not use higher dimensional, in particular 2-dimensional,  (homotopic) systoles to express the entropy of \ $\Sigma$. \ 
We can see at least at least three reasons for this choice.  First,  there is no such thing as a natural ``quantum of area" on \ $\Sigma$, \  in the same sense as there is a natural 
minimum length, the Planck 
length \ $\l_P$. \ A second reason is that we consider space-like surfaces \ $\Sigma$ \ of strictly negative curvature. While surfaces \ ($\Sigma, \mathbf{g}_\Sigma$) \ of curvature 
of either sign are possible in the above argument, we believe that in a semi-classical treatment of stationary black holes, the metric everywhere will  be more uniform, 
at least for surfaces that are of locally maximal area under infinitesimal perturbations. Marginally outer trapped surfaces, which are frequently used in considerations of black hole entropy have such property. 
Such surfaces are used to determine the volume of the interior of a black hole, 
hence, based on conventional statistical mechanical arguments, they  may also be used to determine the entropy of the black hole \cite{CRov, BengJak} in a 
quasi-equilibrium state. Such surfaces, or  any manifolds of 
strictly negative sectional curvature, are aspherical, namely all their higher homotopy groups \ $\pi_i(\Sigma), \ i \geq 2$ \ are trivial, according to the Hadamard-Cartan 
theorem \cite{Gr-book}. Since the higher homotopy groups of \ $\Sigma$ \ vanish, this makes the use of higher
dimensional systoles meaningless. A third reason is far more superficial, but also more pragmatic: far less is known about higher dimensional (homotopy) systoles
than for 1-dimensional ones, and new results for such cases have been very hard to obtain during the last decades. 
So, unless there is overwhelming reason to the contrary, we will rely on results that are currently known, thus use 1-dimensional systoles.\\           

 It is may be worth noticing the similarities between (21) and the definition of the topological entropy \ $h_{top} (\mathcal{M}, \mathbf{g})$ \cite{KatHas}. 
 This is no accident. Without going into any details about the definition an properties of \ $h_{top}$, \  which can be found in \cite{KatHas} for instance, 
 it suffices to notice that 
 \begin{equation} 
      h_{top}(\mathcal{M}, \mathbf{g}) \ \geq \ h_{vol}(\mathcal{M}, \mathbf{g}) 
  \end{equation}
The equality holds in (22) when the metric \ $\mathbf{g}$ \ has no conjugate points.  This is  true when \ $\mathbf{g}$ \  is a metric of strictly negative curvature  
on \ $\Sigma$, \  for instance, as we assume in our case. \\

Another way of seeing why \ $h_{vol}$ \ is a reasonable choice for a function expressing the complexity of \ $\Sigma$ \ is by comparing it with the algebraic entropy of its 
fundamental group \cite{KatHas}. The latter appears, at first sight, to be a better choice for our purposes.  It turns out however, that these two choices give essentially 
equivalent results. To see this, let \ $G$ \ be a discrete finitely generated group and let \ $\Gamma$ \ be a set of its generators. The word length 
$|\gamma |_\Gamma \in \mathbb{N}$ of \ $\gamma\in G$ \ is the length 
of the shortest word though which \ $\gamma$ \ can be expressed in terms of elements of \ $\Gamma$.  \ Let the ball centered at the origin and having 
radius \ $R$ \ in the word metric be  
\begin{equation}      
  B_\Gamma (R) \ = \ \mathrm{card}  \{ \gamma\in G: |\gamma |_\Gamma \leq R \}
\end{equation}
where ``card" stands for the cardinality of the set. Then the algebraic entropy \ $h_{alg}(G, \Gamma)$ \ of \ $G$ \ with respect to \ $\Gamma$ \ is defined as 
\begin{equation}
   h_{alg} (G, \Gamma) \ = \ \lim_{R\rightarrow +\infty} \frac{\log (B_\Gamma (R))}{R} 
\end{equation}
in analogy with (20). A more careful treatment is generally needed in the definition of (24), which however is sufficient for our purposes. 
An upper bound for \ $h_{alg}(G, \Gamma)$ \ is given in terms of the cardinality card($\Gamma$) of \ $\Gamma$ \ by   
\begin{equation}
   h_{alg}(G, \Gamma) \ \leq \ \log (2 \ \mathrm{card}(\Gamma) -1)
\end{equation}
The minimal 
algebraic entropy \ $h_{alg}(G)$ \ of \ $G$ \ is defined to be 
\begin{equation}
       h_{alg}(G) \ = \ \inf_\Gamma h_{alg}(G, \Gamma)
\end{equation}
where the infimum runs over all generating sets \ $\Gamma$ \ of \ $G$. \ The relation between the volume entropy \ $h_{vol}$ \ and the algebraic entropy
\ $h_{alg}$ \ for the spaces of interest to us is as follows. Let \ ($\mathcal{M}, \mathbf{g}$) \ be a closed Riemannian manifold, let \ $G = \pi_1 (\mathcal{M})$ \ 
be its fundamental group having a finite generating set \ $\Gamma$. \ If the norms, induced by the metrics of \ $\Gamma$ \ and \ $\mathcal{M}$ \  satisfy the inequality  
\begin{equation}
       c_1 |\cdot |_\Gamma \ \leq \ | \cdot |_\mathbf{g} \ \leq c_2 |\cdot |_\Gamma
\end{equation}
for positive constants \ $ c_2 \geq c_1 > 0$ \ then 
\begin{equation}
    \frac{1}{c_2} \ h_{alg} (G, \Gamma) \ \leq \ h_{vol}(\mathcal{M}, \mathbf{g}) \ \leq \ \frac{1}{c_1} \  h_{alg} (G, \Gamma)  
\end{equation}
Then (28) guarantees that the algebraic and the volume entropies of such a closed manifold are not too different from 
each other if someone does not really look to distinguish between them in any great detail. In more technical terms, the Lipschitz equivalence of
\  $\mathbf{g}$ \ and the word metric of \ $\Gamma$ \ induces a Lipschitz equivalence between \ $h_{alg}$ \ and \ $h_{vol}$. \  
 Hence, roughy speaking, the results that are obtained by using the 
algebraic entropy of the fundamental group of \ $\mathcal{M}$ \ are the same as the ones obtained by using the volume entropy of \ $\mathcal{M}$.  \\

To proceed, it may be worth recalling Katok's inequality \cite{Katz-book}, which states that any metric \ $\mathbf{g}$ \ on  a closed surface \ $\Sigma$ \ 
having area \ $A (\Sigma)$ \ and negative Euler characteristic \ $\chi(\Sigma)$, \ hence genus \ $g\geq 2$, \ satisfies the inequality   
\begin{equation}
    [h_{vol} (\Sigma, \mathbf{g})]^2 \ \geq \ \frac{2\pi |\chi(\Sigma) |}{A (\Sigma)}
\end{equation}
It is however known, on thermodynamic grounds \cite{Bek1, Bek2, Hawk1} that the entropy of a black hole should be proportional to its area. 
Therefore Katok's inequality (29), which is incidentally also valid for \ $h_{top}$, \   
forces us to consider as the actual, statistical (``Boltzmann/Gibbs/Shannon") entropy of a black hole, not its volume entropy \ 
$h_{vol}(\Sigma, \mathbf{g}_\Sigma)$, \ but instead a multiple of
\begin{equation} 
     \mathcal{S}_{BGS} (\Sigma, \mathbf{g}_\Sigma) \ \equiv \  \frac{1}{h_{vol}^2(\Sigma, \mathbf{g}_\Sigma)}
\end{equation}



\subsection{On entropically-related optimal metrics}

The optimal systolic ratio \  $\mathsf{SR}(\mathcal{M})$ \  of a  manifold \ $\mathcal{M}$ \ is defined as  \cite{KS1}
\begin{equation}
     \mathsf{SR}(\mathcal{M}) \ = \ \sup_\mathbf{g} \mathsf{SR}(\mathcal{M}, \mathbf{g})
\end{equation}
where the supremum is taken over the space of all admissible Riemannian metrics of \ $\mathcal{M}$. \ Since from Katok's inequality (29) we see that 
 \begin{equation}
       h_{vol}(\Sigma, \mathbf{g}) \ \sim \ \frac{1}{\mathrm{Length}}
 \end{equation}
a more appropriate quantitative measure of disorder/complexity of \ $\Sigma$ \ could be the scale-invariant minimum entropy \ $h_{min}$ \  
which is defined as     
\begin{equation}
   h_{min} (\Sigma) \ = \ \inf_\mathbf{g} \ \left\{ h_{vol}(\Sigma, \mathbf{g}) [A(\Sigma, \mathbf{g})]^\frac{1}{2} \right\}
\end{equation}
A relation between the minimal entropy and the optimal systolic ratio of a surface is given by the Katz-Sabourau inequality \cite{KS1}
\begin{equation}
     h_{min}(\Sigma) \ \leq \ - \frac{1}{b L} \log (2a^2 \mathsf{SR} (\Sigma))
\end{equation}
where \ $a,b >0$, \ $4a+b < \frac{1}{2}$ \ and \ $L$ \ stands for the length of the systole of \ $\Sigma$ \ with the optimal metric \ $\mathbf{g}$. \ 
Assuming that the statistical entropy \ $\mathcal{S}_{BGS}$ \ of the black hole is a function of the minimal entropy \ $h_{min}(\Sigma)$, \ we get a rough estimate
for the leading dependence of \ $\mathcal{S}_{BGS}$ \ on the genus \ $g$ \ of \ $\Sigma$ \ by combining (9), (34) as
\begin{equation}  
    \mathcal{S}_{BGS} (\Sigma, \mathbf{g}_\Sigma)  \ \sim \ f(\log g)
\end{equation}
where \ $f$ \ is an appropriate real function, being the inverse square in (30), for instance. 
The form of \ $f$ \ 
depends on the specific identification that one makes between the statistical entropy \ $\mathcal{S}_{BGS}$ \  
and the dynamical entropy of choice of \ $\Sigma$. \ We also see from (9) and (34) that the sub-leading corrections of \ $\mathcal{S}_{BGS}(\Sigma, \mathbf{g}_\Sigma)$ 
\ in terms of the genus are of the form \ \ $f(\log (\log g))$.  \\

We are somewhat skeptical about  proposing as a valid measure of the ``lack of order" / ``complexity" of \ $\Sigma$ \  its minimum entropy \ $h_{min}(\Sigma)$. \
Our attitude is similar toward the
optimal systolic ratio \ $\mathsf{SR}(\Sigma)$, \ and thus about using the Katz-Sabourau inequality (34) to get a lower bound for the statistical entropy of \ $\Sigma$, \
taking into account an identification of the inverse, in the spirit of (30). Our reservations arise from the fact that it is not clear to us that the states violating the classical
energy inequalities make a statistically sufficient contribution  to allow the semi-classical metric $\mathbf{g}$ of space-time, which induces the metric of \ $\Sigma$,
 \  to explore the whole space of metrics of \ $\Sigma$. \ 
Stability of the semi-classical solution under small perturbations necessarily confines the allowed space of metrics to a neighborhood  of the classical one \ 
$\mathbf{g}$. \ Hence, it is not apparent that the supremum in (31) and the infimum in (33) can  physically even be approached. \\   


\subsection{The Lusternik-Schnirelmann and systolic categories}

Another motivation for using a function of systoles  as a measure of complexity, hence as a potential measure of the entropy of a black hole, comes through 
the relation between the Lusternik-Schnirelmann (for a relatively recent survey see \cite{LuSc}) and the systolic categories of \ $\Sigma$. \  
It should be mentioned at this point that the term ``category" in this subsection has 
nothing to do with ``category theory" which is an increasingly popular field used in many branches of Mathematics, and even of Physics. \\   

The Lusternik-Schnirelmann category \cite{LuSc} \ $\mathsf{cat}_{LS} (\mathfrak{X})$ \ of a topological space \ $\mathfrak{X}$ \ is the least \ $k\in\mathbb{N}$ \  
such that there is an open covering \  $\mathcal{U}_i, \ i=1, \ldots, k+1$ \  such that each \ $\mathcal{U}_i$ \ is contractible in \ $\mathfrak{X}$. \ 
If no such \ $k\in\mathbb{N}$ \ exists, then we set \ $\mathsf{cat}(\mathfrak{X}) = \infty$. \ The Lusternik-Schnirelmann category gives a quantitative measure of
the level of complexity of a space by expressing how many contractible sets one needs to cover it  which, from a homotopic viewpoint, are the simplest 
possible sets. From a physical viewpoint, these $\mathcal{U}_i$ are the ``quanta" of the the homotopy characterization of \ $\Sigma$. \ 
For \ $\mathcal{M} = \mathbb{S}^2$ \ we easily see  that \ $\mathsf{cat}_{LS}(\mathbb{S}^2) = 1$ \ and for all higher genus \ $g\geq 1$ \ surfaces \ 
$\Sigma_g$ \ that \ $\mathsf{cat}_{LS} (\Sigma_g) = 2$.  \\

The systolic category was introduced in \cite{KatzR} as a Riemannian analogue, par excellence, of the Lusternik-Schnirelmann category. Without going into too many 
details in its definition, as they can be found in \cite{Katz-book} and would take us too far afield, one defines the systolic category \ $\mathsf{cat}_{sys} (\Sigma, \mathbf{g})$ \ 
of a 2-dimensional surface \ $\Sigma$ \  to be the largest integer \ $d$ \ such that  
\begin{equation}
    [Sys (\Sigma, \mathbf{g})]^2 \ \leq \  \mathrm{C(\Sigma)} \ A(\Sigma, \mathbf{g})
\end{equation}
 for all Riemannian metrics \ $\mathbf{g}$ \ on \ $\Sigma$, \  where \ $C$ \ is a positive constant which depends on \ $\Sigma$ \ but not on its metric \ $\mathbf{g}$. \ 
 Then for all  2-dimensional surfaces \ $\Sigma$, \ one can see that
\begin{equation}
     \mathsf{cat}_{LS} (\Sigma) \ = \ \mathsf{cat}_{sys} (\Sigma)
\end{equation}
As a result, the homotopic Lusternik-Schnirelmann viewpoint and the Riemannian viewpoint through systoles, expressed via the corresponding categories, 
give the same result  for the complexity of all surfaces \ $\Sigma$ \  needed  in our considerations.\\


\newpage

\section{Conclusion and outlook}

In this work, we have presented a patchwork of statements, rather than one unified argument,  purporting to support the idea that 
it may not be totally unreasonable to ascribe the entropy of black holes in $(3+1)$-dimensional spacetimes close to quasi-equilibrium  
to the homotopy systolic properties of the space-like sections \ $\Sigma$ \ of their horizons.  We have assumed throughout the present work, maximal 
violation of the classical energy inequalities, and of the resulting topological censorship. This allows the existence of horizons whose space-like sections 
are surfaces \ $\Sigma$ \  of genus  \ $g$, \ where \ $g$ \ can become very large. We have used the volume entropy of \ $\Sigma$ \ as the basic quantity 
in mesoscopically encoding the ``lack of order"/``complexity" that the entropy of the black hole should express. 
The relations with other dynamically-motivated entropies, systoles and the Lusternik-Schnirelmann category are also 
brought forth in presenting our viewpoint.\\      
 
The advantage of using systolic inequalities is that they are curvature-free. This way, we bypass our inability to control the effect of quantum mechanical states that 
violate the classical energy inequalities. Their curvature-free property makes such inequalities quite robust, but at the same time this insensitivity pays the price 
of not being able to detect the finer properties of the stress-energy tensor, even at the semi-classical level.  
One could  have equally well used homology systoles, instead of homotopic ones,  but they seem to be less robust, or less  natural, in our opinion, 
as a measure of the complexity of  \ $\Sigma$ \  giving rise to the entropy of black holes \cite{Katz-book}. \\

In a companion paper to the present one \cite{NK1} we have investigated how the present arguments need to be modified for the well-known case 
of \ $\Sigma = \mathbb{S}^2$. \  Since \ $\pi_1(\mathbb{S}^2)$ \ is trivial, one has to replace the systoles in probing such horizons with another geometric feature 
of these 2-spheres. In \cite{NK1} we argued that the injectivity radius and the related embolic inequalities may be the appropriate analogues of the systoles and the 
systolic inequalities discussed above, for the case of \  $\Sigma = \mathbb{S}^2$. \\  


               \vspace{10mm}

\noindent{\normalsize\bf Acknowledgement:}  \ \ We are  grateful to Professor Anastasios Bountis  for his encouragement and support without which this work
would have never been possible.  \\  



\end{document}